\documentclass[aps,pra,twocolumn,superscriptaddress]{revtex4-1}
\usepackage{graphicx}
\usepackage{amssymb}
\usepackage{amsmath}
\usepackage{bm}
\usepackage{xcolor}
 \usepackage{verbatim}

\begin{document}

\title{Vortex shedding frequency of a moving obstacle in a Bose-Einstein condensate}

\author{Younghoon Lim}
\affiliation{Department of Physics and Astronomy, Seoul National University, Seoul 08826, Korea}
\affiliation{Center for Correlated Electron Systems, Institute for Basic Science, Seoul 08826, Korea}

\author{Yangheon Lee}
\affiliation{Department of Physics and Astronomy, Seoul National University, Seoul 08826, Korea}
\affiliation{Center for Correlated Electron Systems, Institute for Basic Science, Seoul 08826, Korea}

\author{Junhong Goo}
\affiliation{Department of Physics and Astronomy, Seoul National University, Seoul 08826, Korea}

\author{Dalmin Bae}
\affiliation{Department of Physics and Astronomy, Seoul National University, Seoul 08826, Korea}
\affiliation{Center for Correlated Electron Systems, Institute for Basic Science, Seoul 08826, Korea}

\author{Yong-il Shin}
\email{yishin@snu.ac.kr}
\affiliation{Department of Physics and Astronomy, Seoul National University, Seoul 08826, Korea}
\affiliation{Center for Correlated Electron Systems, Institute for Basic Science, Seoul 08826, Korea}
\affiliation{Institute of Applied Physics, Seoul National University, Seoul 08826, Korea}

\begin{abstract}
We experimentally investigate the periodic vortex shedding dynamics in a highly oblate Bose-Einstein condensate using a moving penetrable Gaussian obstacle. The shedding frequency $f_v$ is measured as a function of the obstacle velocity $v$ and characterized by a linear relationship of $f_v=a(v-v_c)$ with $v_c$ being the critical velocity. The proportionality constant $a$ is linearly decreased with a decrease in the obstacle strength, whereas $v_c$ approaches the speed of sound. When the obstacle size increases, both $a$ and $v_c$ are decreased. The critical vortex shedding is further investigated for an oscillating obstacle and found to be consistent with the measured $f_v$. When the obstacle's maximum velocity exceeds $v_c$ but its oscillation amplitude is not large enough to create a vortex dipole, we observe that vortices are generated in the low-density boundary region of the trapped condensate, which is attributed to the phonon emission from the oscillating obstacle. Finally, we discuss a possible asymptotic association of $a$ with the Strouhal number in the context of universal shedding dynamics of a superfluid.

\end{abstract}

%\pacs{67.85.De, 03.75.Lm, 03.75.Kk}

\maketitle

\section{Introduction}

In fluid dynamics, the flow pattern behind a moving obstacle is a classic topic of fundamental and practical interest~\cite{LLtext}. It is known that the wake pattern evolves from laminar to turbulent as the obstacle's velocity increases, and the transition is well characterized by the Reynolds number that is defined as $\textrm{Re}=v D/\nu$ with $v$ and $D$ being the obstacle's velocity and diameter, respectively, and $\nu$ the fluid's kinematic viscosity. $\textrm{Re}$ represents the ratio of inertial force to viscous force in the fluid and explains that the fluid's response becomes turbulent when the inertial force associated with the moving obstacle exceeds the `stickiness' of the fluid. A noticeable observation in the wake pattern transition is the periodic generation of vortices with alternating circulations for a wide range of $50 < \textrm{Re} < 10^5$~\cite{Lienhard}, which is known as the von Kármán vortex street. Intriguingly, the Strouhal number, which is a dimensionless quantity defined as $\textrm{St}=f_v D/v$ with $f_v$ being the vortex shedding frequency, is nearly constant at $\approx 0.2$ in the periodic shedding regime. This periodic vortex shedding is a universal characteristic of the classical viscous fluid.

Superfluidity, the absence of friction in the fluid, presents an interesting situation for the wake pattern problem, where energy dissipation occurs only above a certain critical velocity, and vortices should have quantized circulations, thus called quantum vortices. A central question is how the wake pattern is formed and evolves in a superfluid, particularly in comparison to the universal behavior of classical fluids. Atomic superfluid gases provide an excellent experimental platform to address this question, where one can apply local perturbations to samples using various optical means and also directly image quantum vortices. Since the first realization of atomic Bose-Einstein condensates (BECs)~\cite{Anderson,Davis}, the critical vortex dynamics have been widely investigated theoretically~\cite{Frisch92,Jackson98,Winiecki99,Winiecki00,Sasaki10,Aioi11,Reeves15} and experimentally~\cite{Raman99,Inouye01,Neely10,Kwon15-cv,Kwon15-vd,Kwon16-vk,Park18}. In the experiment, a moving obstacle was formed by focusing a laser beam and the vortex generation was demonstrated via matter wave interference~\cite{Inouye01} and by observing the density-depleted cores of created quantum vortices~\cite{Neely10}. Highly oblate samples have facilitated systematic studies on the vortex shedding dynamics~\cite{Neely10,Kwon15-cv,Kwon15-vd,Kwon16-vk,Park18}, where vortex lines tend to be aligned to the tight confining direction so that the vortex dynamics is effectively two-dimensional (2D). The critical velocity $v_c$ for vortex shedding was measured for various obstacle parameters, supporting the theoretical description based on the local Landau criterion \cite{Kwon15-cv}. Moreover, the regular shedding of vortex clusters consisting of like-sign vortices, a quantum version of the von Kármán vortex street~\cite{Sasaki10}, was experimentally demonstrated~\cite{Kwon16-vk}. In particular, the Strouhal number was found to exhibit saturation behavior with increasing $v$ and its saturation value was estimated to be about 0.2, remarkably similar to that of the classical fluid. Reeves {\it et al.}~\cite{Reeves15} numerically showed that the shedding pattern in a 2D superflow exhibits dynamical similarity for various obstacle diameters, from which they suggested a modified Reynolds number for the universal superfluid dynamics, employed to the discussion of turbulence in superfluid $^{4}\textrm{He}$~\cite{Schoepe15}.

In this paper, we present an experimental study of the periodic vortex shedding dynamics in a highly oblate BEC for a {\it penetrable} obstacle. Being penetrable means that the obstacle has a potential barrier lower than the condensate's chemical potential so that the superfluid can penetrate the obstacle. This situation is qualitatively different from the typical hard cylinder case, in which a zero-density region is not induced by the obstacle in the superfluid and therefore, vortices cannot be created individually but in a form of dipoles consisting of two vortices of opposite circulations. Periodic shedding of vortex dipoles from a moving penetrable obstacle was observed in previous experiments~\cite{Kwon15-vd}. In this work, motivated by the characteristic behavior of $\textrm{St}$ observed for an impenetrable obstacle~\cite{Kwon16-vk}, we investigate the vortex-dipole shedding frequency $f_v$ of a penetrable obstacle and its dependence on the obstacle's parameters such as potential barrier height and size. We measure  the shedding frequency $f_v$ as a function of the obstacle velocity $v$, and characterize it by a linear relationship of $f_v=a(v-v_c)$. We find that as the obstacle strength decreases, the proportionality constant $a$ is linearly decreased, whereas $v_c$ approaches the speed of sound $c_s$. When the obstacle size increases, both $a$ and $v_c$ are decreased. In addition, we investigate the vortex shedding dynamics from an oscillating obstacle, and observe its critical behavior consistent with the measured $f_v$. Particularly, when the obstacle moves faster than $v_c$ but the oscillation amplitude is not sufficient to create a vortex dipole, we observe that vortices are generated in the low-density boundary region of the trapped condensate, which we attribute to the phonon emission from the oscillating obstacle. Finally, we discuss a possible asymptotic association of $a$ with the Strouhal number as the shedding dynamics evolves into the impenetrable regime as the obstacle strength increases. This work provides comprehensive information on the periodic shedding dynamics in a BEC for a penetrable obstacle, which would be beneficial to establish the phenomenological understanding of universal superfluid dynamics.

\section{Periodic vortex shedding model}

In this section, we briefly describe our model of the periodic vortex shedding from a moving obstacle in a superfluid. We consider a penetrable obstacle moving with velocity $v$ in a BEC (figure 1(a)), where $V_0$ and $\mu$ denote the obstacle's potential height and the condensate's chemical potential, respectively, and $V_0<\mu$. When the obstacle moves faster than a critical velocity $v_c$, energy is transferred into the condensate in a way of changing the superfluid velocity field near the obstacle. Here we assume that the energy transfer rate is given by $P=\Gamma(v-v_c)$, where $\Gamma$ is a proportionality constant with dimensions of force~\cite{Kwon15-vd,Park18}. This means the generation of a drag force $F=P/v\approx \Gamma (\frac{v}{v_c}-1)$ for $v-v_c\ll v_c$, which is supported by previous numerical studies~\cite{Frisch92,Winiecki99,Winiecki00}. Then, the energy is accumulated as $E=Pt_s=\Gamma t_s(v-v_c)$ for a duration $t_s$, and when $E$ exceeds a certain energy cost $E_v$ for creating a vortex dipole, it will dissipate via vortex generation. If the energy accumulation is not significantly affected by the velocity field of the created vortex dipoles~\cite{Kwon15-vd,Kadokura14}, the vortex generation process will be repeated during the obstacle translation, leading to the periodic shedding of the vortex dipoles (figure 1(b)). Then, the shedding frequency $f_v$ is given by $f_v=P/E_v=a(v-v_c)$, where $a=\Gamma/E_v$ with dimensions of an inverse of length. In our model, the periodic vortex shedding is characterized by the two quantities, $v_c$ and $a$, which would be determined by the obstacle properties such as strength ($V_0$), size, and shape.

%%%%% figure 1

\begin{figure}[t]
 \includegraphics[width=82mm]{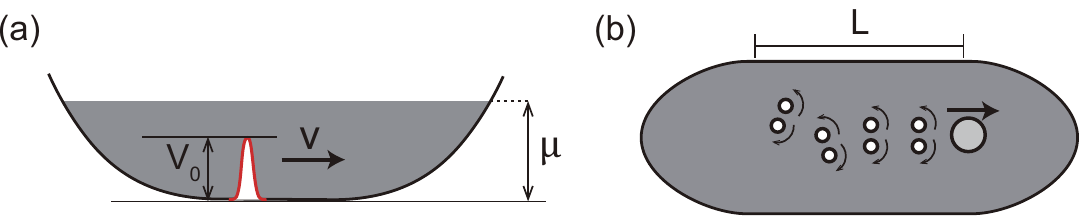}
 \centering
  \caption{Periodic vortex shedding in a Bose-Einstein condensate (BEC). (a) An obstacle moves with a velocity $v$ in a BEC confined in a trapping potential. The obstacle is penetrable with a potential height $V_0$ less than the chemical potential $\mu$ of the condensate. (b) When the obstacle moves faster than a certain critical velocity $v_c$, vortex dipoles are periodically shed from the obstacle. $L$ denotes the traverse distance of the obstacle in the trapped condensate.}
\end{figure}

%%%%%%%%%%%

For a reliable measurement of $f_v$, the sweeping distance $L$ of the obstacle should be long enough to have multiple shedding events. From the linear relationship $f_v=a(v-v_c)$, the required condition of $L$ to obtain vortex dipoles more than $N_d$ is given by 
\begin{equation}
    L > L_{N_d} = N_d \frac{v}{f_v}= \frac{N_d v}{a(v-v_c)}.
\end{equation}
Since $L_{N_d}$ diverges for $v\rightarrow v_c$, it would be difficult to measure $f_v$ near the critical velocity, which is the case in typical experiments using a trapped sample with finite spatial extent. If we re-express Eq.~(1) for the minimum obstacle's velocity to generate vortex dipoles for the given $L$, i.e., $N_d\geq 1$, it gives
\begin{equation}
     v \geq v_{c,L} = \frac{aL}{aL-1}v_c.
\end{equation}
This means that in the extreme case of $aL<1$, vortex shedding itself would be improbable for the experimental demonstration. Therefore, it is clear that having a large-area sample that allows a long $L$ will be highly beneficial to investigate the shedding dynamics and particularly, to explore its universality for a wide range of the parameter space. 

%%%%% figure 2

\begin{figure*}[t]
 \includegraphics[width=125mm]{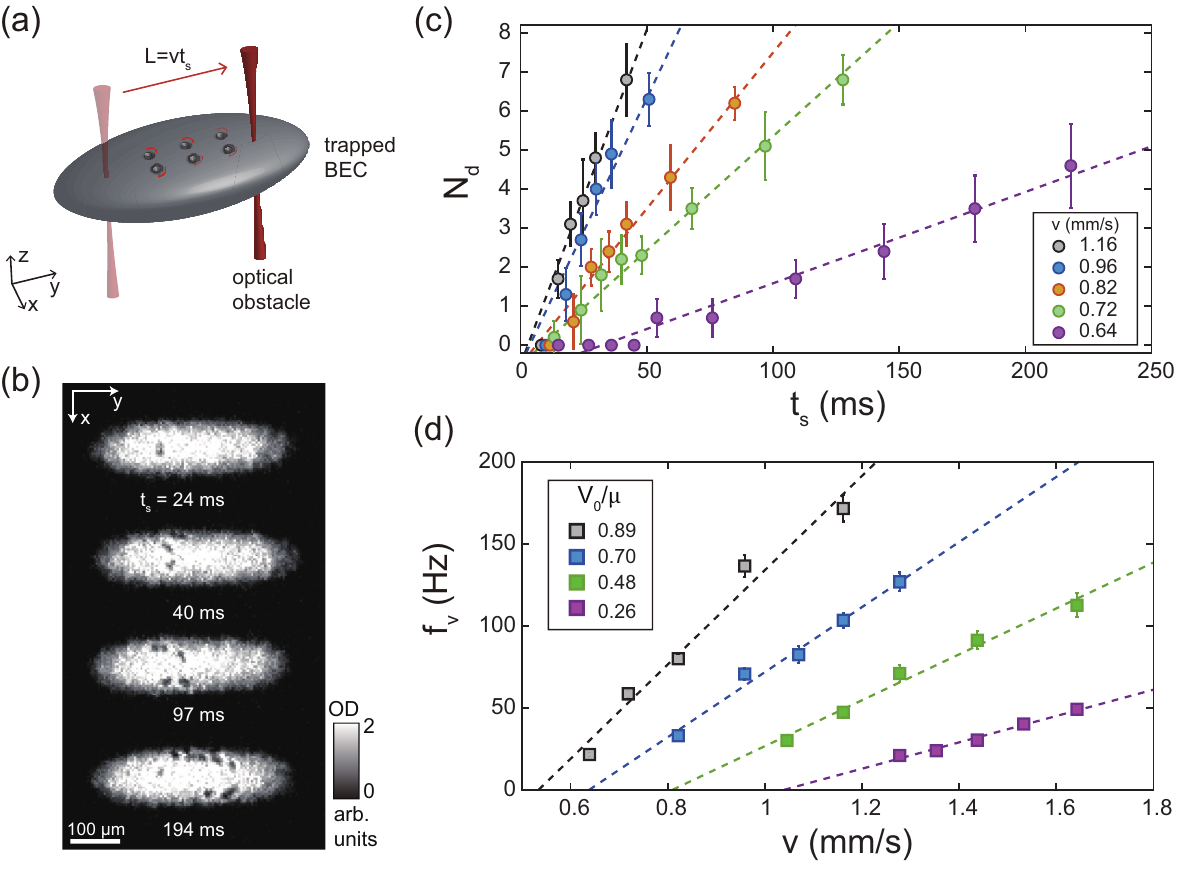}
\centering
  \caption{Measurement of the vortex shedding frequency $f_v$. (a) Schematic of the shedding experiment. A penetrable optical obstacle, created by focusing a repulsive Gaussian laser beam, is linearly translated across a highly oblate and elongated BEC with a constant velocity $v$ for a time $t_s$. (b) Absorption images of the BEC for various sweeping times $t_s$ with $v=0.72~\textrm{mm/s}$. The images were obtained after a 40.9 ms time-of-flight to detect created vortex dipoles by their expanded, density-depleted cores. The obstacle strength and size were $V_0/\mu=0.89$ and $\sigma=7.1~\mu\textrm{m}$, respectively. $\sigma$ denotes the $1/e^2$ radius of the obstacle potential. (c) Vortex dipole number $N_d$ as a function of $t_s$ for various $v$ with $V_0/\mu=0.89$ and $\sigma=7.1~\mu\textrm{m}$. Each data point was obtained from ten measurements of the same experiment, and its error bar represents their standard deviation. The vortex shedding frequency $f_v$ is determined by fitting a linear function of $N_d=f_v t_s -\frac{1}{2}$ (dashed line) to the data. (d) $f_v$ as a function of $v$ for various $V_0/\mu$ ($\sigma=7.1~\mu\textrm{m}$). The dashed line indicates a linear fit of $f_v=a(v-v_c)$ to the data, from which the critical velocity $v_c$ and the proportionality constant $a$ are determined.}
\end{figure*}

%%%%%%%%%%%

\section{Experiment}

Our experiment is performed with a BEC of $^{87}$Rb in an optical dipole trap (ODT). To prepare a sample with large area, we use a clipped Gaussian ODT as described in Ref.~\cite{Lim21}, where a 1064-nm Gaussian laser beam is symmetrically truncated by a horizontal slit and vertically focused through a cylindrical lens to form a highly oblate ODT. Due to the clipping, the ODT is not only elongated along the beam propagation direction but also flattened in the center region. In a typical sample condition, the condensate contains about 9.1$\times10^6$ atoms with Thomas-Fermi radii of $(R_x, R_y, R_z) \approx (61, 239, 2.7)~\mu\textrm{m}$. The condensate fraction of the sample is over 80$\%$. The chemical potential of the condensate is estimated to be $\mu \approx k_B\times42.9~\textrm{nK}$, where $k_B$ is the Boltzmann constant, and the healing length is $\xi=\hbar/\sqrt{2m\mu}\approx0.26~\mu\textrm{m}$, where  $\hbar$ is the reduced Planck constant and $m$ is the atomic mass. At the center of the highly oblate condensate, the speed of sound is given by $c_s=\sqrt{\frac{2\mu}{3m}}\approx1.65~\textrm{mm/s}$~\cite{Stringary98, Kim20}. By virtue of the elongated and flattened geometry of the clipped-Gaussian ODT, the condensate column density is uniform within 10$\%$ near 1000$\xi$ along the long axis. In the sample preparation, fast cooling would result in spontaneous vortex formation during the phase transition~\cite{Goo21}, so we adjusted the cooling curve of the ODT depth to suppress the average vortex number to less than one.

A repulsive optical obstacle is realized by tightly focusing a 532~nm Gaussian laser beam propagating along the $z$ axis, which is perpendicular to the condensate plane. The obstacle width $\sigma$ denotes the $1/e^2$ radius of the obstacle potential and was calibrated from {\it in-situ} images of the BEC penetrated by the repulsive beam for various obstacle strengths, taking into account the imaging resolution. At the focal plane of the laser beam, $\sigma= 7.1(7)~\mu\textrm{m}\approx 27\xi$. In our experiment, the obstacle width is changed simply by defocusing the laser beam at the condensate plane. The Rayleigh length of the laser beam is considerably longer than the condensate thickness and thus, we ignore the divergence of the laser beam as it penetrates the condensate. The position of the obstacle is controlled by a piezo-actuated mirror.

A schematic of the shedding experiment is illustrated in figure 2(a). 
The obstacle beam is initially located at the left ($-y$) side from the condensate center and its intensity is adiabatically switched on to a target value of $V_0/\mu$ for 100~$\textrm{ms}$ to prevent unwanted perturbations to the sample. After an additional hold time of 20~$\textrm{ms}$, the obstacle is linearly translated along the $y$ axis to traverse the center region of the BEC with a constant velocity $v$ for a sweeping duration $t_s$. Then, the obstacle beam is linearly turned off for 200~$\textrm{ms}$ to ensure that the vortex measurement is not affected by the switch-off process. The number of generated vortex dipoles, $N_d$, is measured by taking an absorption image of the condensate after a 40.9-ms time-of-flight and counting the density-depleted holes of the merged cores of vortex dipoles (figure 2(b)). In our imaging, individual vortices that might be created in the sample preparation are distinguishable from vortex dipoles generated by the obstacle due to their different shape of density-depleted cores~\cite{Kwon-rlx}, so not likely to be counted in our $N_d$ measurement.

Our sample provides the maximum distance of the obstacle translation over $200~\mu\textrm{m}$ without significant density variations. The typical initial position of the obstacle is set to be approximately $70~\mu\textrm{m}$ away from the center of the sample. However, when the value of $a$ is excessively low to obtain a sufficient $N_d$, the initial position is moved further away from the center by $\approx 30~\mu\textrm{m}$, which was the case for the two lowest values of $V_0/\mu<0.4$ in our experiment.

\section{Results and Discussions}

\subsection{Determination of shedding frequency}

Figure 2(c) shows the measurement results of $N_d$ as a function of the sweeping time $t_s$ for various velocities $v$ with $V_0/\mu\approx 0.9$ and $\sigma/\xi\approx 27$. According to the periodic shedding model, the vortex dipoles are expected to be created from the moving obstacle in a periodic manner, which would result in a periodic stepwise increase of $N_d$ with increasing $t_s$~\cite{Kwon15-vd}. In our experiment, such stepping behavior was not clearly observed, probably due to the residual motion of the trapped sample as well as the intrinsic stochasticity of the vortex shedding dynamics. Nevertheless, we observe that $N_d$ exhibits a linear dependence on $t_s$, which is consistent with the periodic shedding model in an averaging sense. For the measurement of $f_v$, we increase the scan range of $t_s$ to obtain the mean number of $N_d$ over 4 and determine $f_v$ from a linear fit of $N_d = f_v t_s-\frac{1}{2}$ to the data. 

In figure 2(d), we display the measured $f_v$ as a function of $v$ for various obstacle strengths. It is clearly observed that the $v$-dependence of $f_v$ is well described by the model curve of $f_v=a(v-v_c)$ over the whole range of $V_0/\mu$ in our experiment. From a model curve fit to the experimental data, we determine the critical velocity $v_c$ and the proportionality constant $a$ for each obstacle condition. Our experimental observations confirm that the vortex shedding from a penetrable obstacle can be parametrized by the two quantities $\{v_c, a\}$ as expected from our periodic shedding model. 

%\subsection{Slow superflow defection}

\begin{figure}[t]
 \includegraphics[width=84mm]{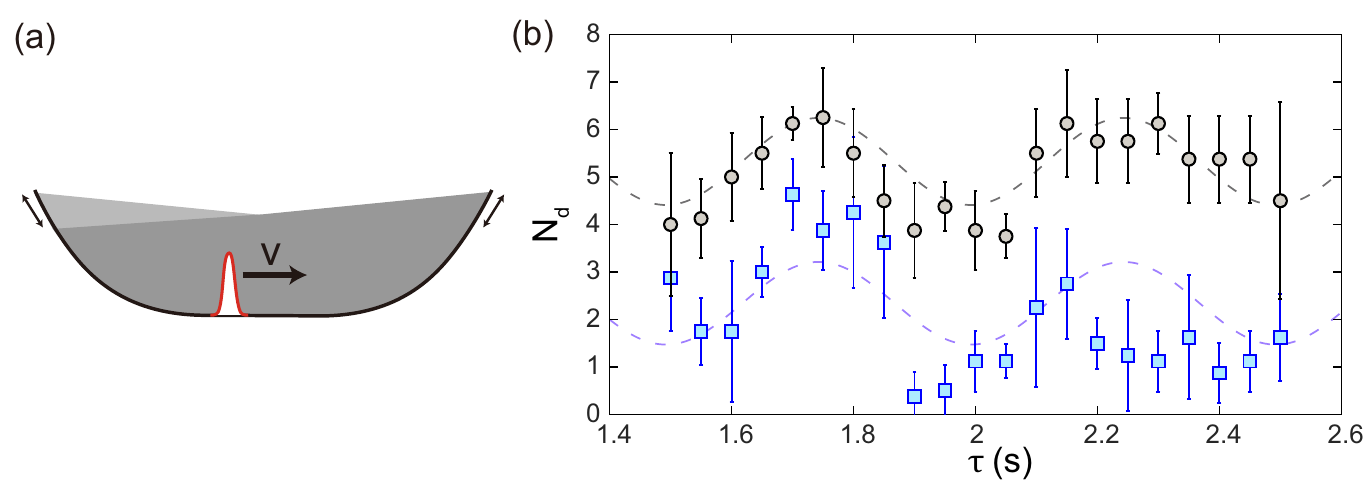}
 \centering
 \caption{Effect of BEC motion on the vortex shedding. (a) Illustration of dipole oscillations of the trapped condensate. Such condensate motions affect the relative velocity of the moving obstacle and consequently, the vortex shedding frequency. (b) Variations of the vortex number $N_d$ for different hold times, $\tau$, taken before starting the obstacle sweeping. The BEC was prepared to have small dipole oscillations. $V_0/\mu=0.5$, $\sigma/\xi=27$, $v-v_c=0.3~\textrm{mm/s}$ (black circle) and $0.15~\textrm{mm/s}$ (blue square), and $t_s=125~\textrm{ms}$. Each data point was obtained from ten measurements of the same experiment, and its error bar indicates their standard deviation. The dashed lines are guides to the eyes.}
\end{figure}

Before presenting the detailed study of the dependence of $\{v_c, a\}$ on the obstacle parameters, we note that preparation of a stationary sample is necessary to measure $f_v$ precisely. When the condensate moves, for example, with a velocity $u$ along the sweeping direction of the obstacle, the relative velocity of the obstacle to the condensate is changed to $v-u$, and the shedding frequency would be shifted to $f_v'=a(v-u-v_c)=f_v-au$, resulting in a change of the number of created vortex dipole by $\Delta N_d=-a \bar{u} t_s$, where $\bar{u}$ denotes the average condensate velocity over the sweeping duration of $t_s$. In our experiment, we observed that when the dipole motion of the BEC was not completely damped down in the ODT, $N_d$ showed resultant oscillations for varying the hold time taken before initiating the obstacle sweeping. Figure 3 shows a couple of examples of such $N_d$ measurements for two different sweeping velocities and $t_s=125~\textrm{ms}$. From the value of $a=154~\textrm{mm}^{-1}$, which was measured for the given obstacle condition of $V_0/\mu\approx0.5$ and $\sigma/\xi\approx27$ (figure 4(a)), the oscillation amplitude of $\bar{u}$ is estimated to be $\approx0.06~\textrm{mm/s}$. This corresponds to the motional energy of atoms as low as $\approx k_B\times 40~\textrm{pK}$, strikingly demonstrating the high velocity sensitivity of the vortex shedding process.

\subsection{Obstacle parameter dependence}

%%%%% figure 3

\begin{figure*}[t]
 \includegraphics[width=140mm]{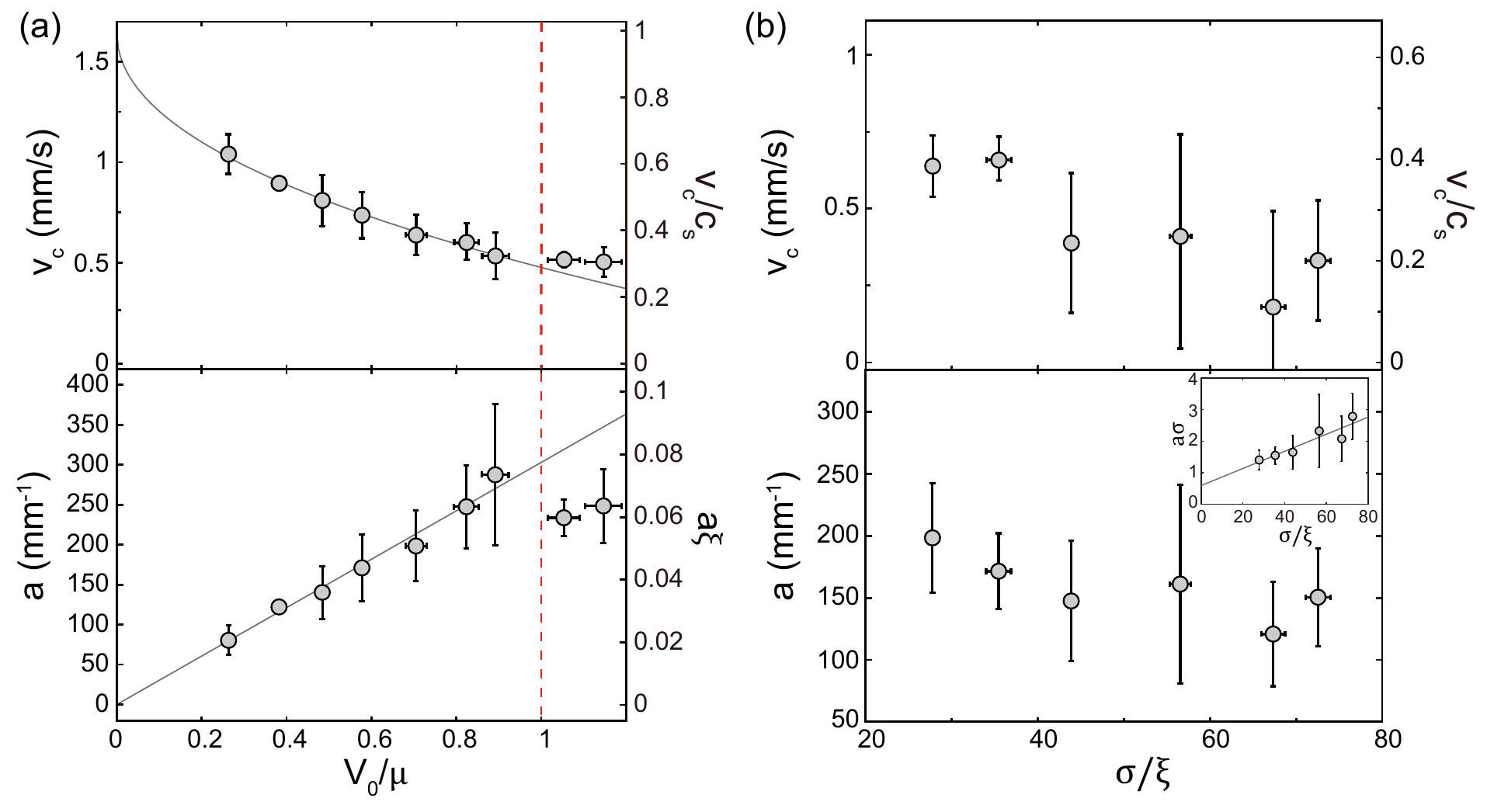}
 \centering
  \caption{Critical velocity $v_c$ and proportionality constant $a$ as functions of (a) the obstacle strength $V_0/\mu$ and (b) the obstacle size $\sigma/\xi$. $\xi$ denotes the healing length of the condensate at its peak density. In (a), $\sigma/\xi= 27$ and in (b), $V_0/\mu=0.70$. The y error bars indicate one standard errors of the fitting (figure~2(a)) and the unseen error bars are hidden by the markers. The solid gray lines in (a) show the curves of $v_c/c_s=1-k (V_0/\mu)^\alpha$ with $k=0.71$ and $\alpha=0.47$ in upper panel and $a= a_{c} (V_0/\mu)$ with $a_c=303~\textrm{mm}^{-1}$ in lower panel. The boundary of the penetrable regime at $V_0/\mu = 1$ is indicated by a red dashed line. The inset in (b) shows the same data of the corresponding panel and the solid line is a linear fit of the data.}
\end{figure*}

%%%%%%%%%%%

In figure~4(a), we first present the measurement results of $\{v_c, a\}$ for our smallest obstacle with $\sigma/\xi\approx 27$ as a function of the normalized obstacle strength $V_0/\mu$. The critical velocity $v_c$ is increased as $V_0/\mu$ decreases. The increase of $v_c$ for weak obstacles is generally understood as a result of the increase of the local density in the obstacle region, i.e., when the obstacle becomes weakened, the atom density of the condensate at the obstacle is less depleted, leading to an increase of the local speed of sound and consequently, the critical velocity according to the Landau criterion~\cite{Kwon15-cv}. In this understanding, when the obstacle strength approaches zero, $v_c$ is expected to reach the speed of sound for the trapped BEC, so we further characterize the dependence of $v_c$ on $V_0/\mu$ by fitting a function of $v_c/c_s=1-k(V_0/\mu)^\alpha$ to the data, yielding $k=0.71(2)$ and $\alpha=0.47(5)$. Meanwhile, we observe that the proportionality constant $a$ decreases linearly with decreasing $V_0/\mu$. Considering $a$ as an indicator of the drag force from the aforementioned discussion of the periodic shedding model, the decreasing $a$ can be interpreted as an alleviation of the drag on the superfluid and it seems reasonable to have $a=0$ for $V_0/\mu\rightarrow 0$ owing to the disappearance of the obstacle. The measurement results for $V_0/\mu<1$ is found to be well described by a linear function fit of $a=a_c (V_0/\mu)$ with $a_c=303(15)~\textrm{mm}^{-1}$ ($a_c\xi\approx 0.08$).

We also examine how $v_c$ and $a$ change as the obstacle evolves from penetrable to impenetrable by extending the scan range of $V_0/\mu$ slightly over unity. This region is interesting because the shedding mechanism might be qualitatively altered due to the possibility of generating individual vortices~\cite{Kwon15-vd,Kwon16-vk}. We observe that $v_c$ remains almost the same at $\approx 0.3 c_s$, whereas $a$ is slightly decreased from its peak value $\approx a_c$. The prosaic response of $v_c$ is in contrast with the previous result of Ref.~\cite{Kwon15-cv}, where the critical velocity was observed to exhibit a dip at $V_0/\mu\approx 1$ in the penetrable-to-impenetrable transition. We attribute the discrepancy to the different methods used to measure the critical velocity. In the previous work, the critical velocity was determined as the minimum obstacle velocity for generating vortices for a given sweeping distance, i.e., $v_{c,L}$ and therefore, as discussed in Sec.~2, it would overestimate the true $v_c$, probably, by $\Delta v_c=v_{c,L}-v_c=\frac{1}{aL-1}v_c$ and the dip structure of $v_{c,L}$ might result from the hump of $a$ which we observed at $V_0/\mu\approx 1$. Note that in the current work, the critical velocity is determined from the $f_v$ measurements and thus, free from the systematic effect due to the finite sweeping distance. The reduction of $a$ in the impenetrable regime will be further discussed in section~4.4.

%%%%% figure 4

\begin{figure*}[t]
 \includegraphics[width=140mm]{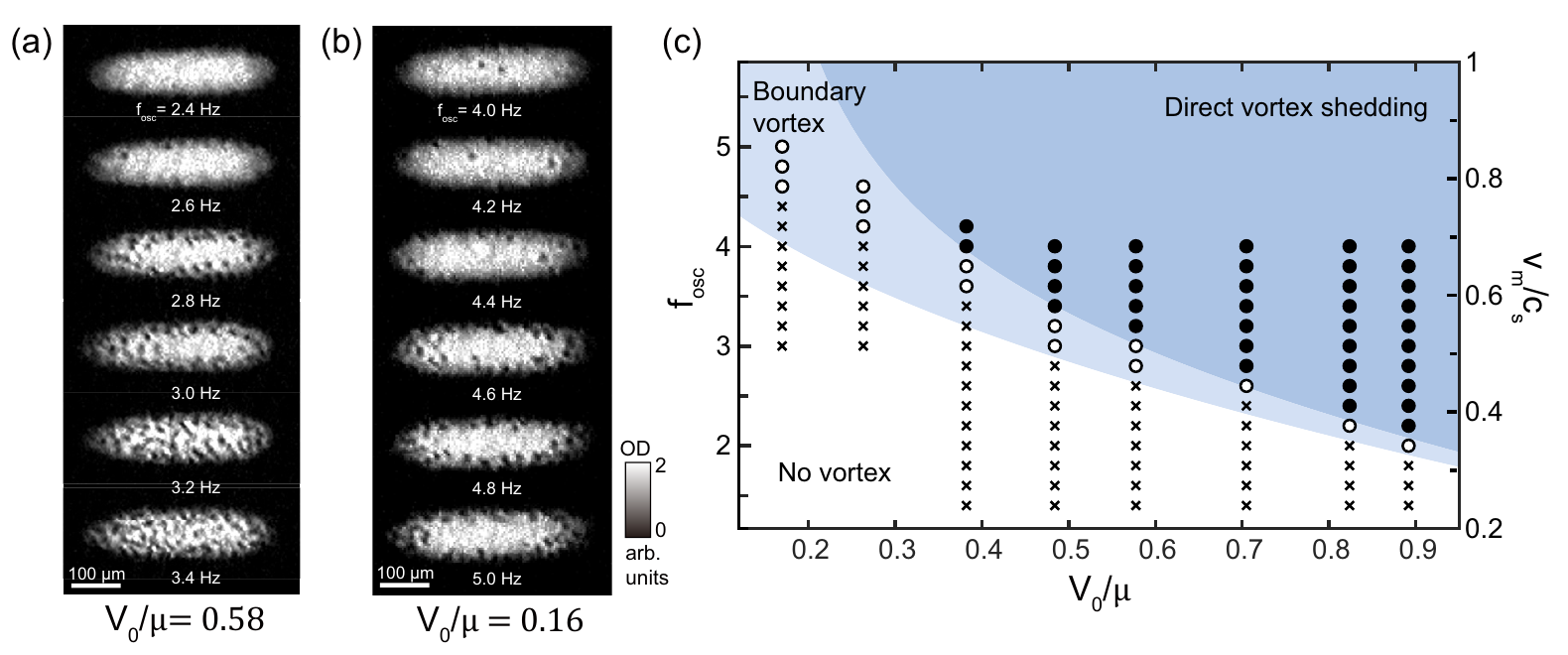}
 \centering
  \caption{Vortex generation via oscillating penetrable obstacle in the trapped BEC. Images of BEC with various oscillation frequencies $f_{osc}$ for (a) $V_0/\mu=0.58$ and (b) 0.16. As $f_{osc}$ increases, vortices first appear in the boundary region of the trapped condensate and then, populates in the center region for higher $f_{osc}$. (c) Various vortex generation regimes in the plane of $V_0/\mu$ and $f_{osc}$. The marker shape indicates the experimental observation at its position; cross: no vortex generation from the obstacle, open circle: vortex only at the boundary of BEC, solid circle: direct vortex shedding from the obstacle. Different shaded areas indicate the three dissipation regimes estimated from the measurement results of $\{v_c, a\}$ in figure~4(a) (see text for detail).}
\end{figure*}

%%%%%%%%%%%

Next, we investigate the dependence of $\{v_c, a\}$ on the obstacle size. Figure~3(b) shows the measurement results as a function of $\sigma/\xi$ for a fixed obstacle strength $V_0/\mu\approx 0.7$. For $25<\sigma/\xi<75$, $v_c$ is decreased from $\approx 0.4 c_s$ to $\approx 0.2 c_s$ with $\sigma/\xi$ increasing, and $a$ is also reduced by a factor of, similarly, about 2. The $\sigma/\xi$ dependence of $v_c$ implies that the critical shedding is not only determined by the local density at the tip of the obstacle but also the density curvature around it, which seems reasonable due to the fact that vortex dipoles are objects with a finite-size structure~\cite{Kwon15-cv}. Currently, however, we have no plausible explanation on how both $v_c$ and $a$ are decreased for a larger obstacle. When vortices are readily generated with a smaller critical velocity, it might be generally expected that the vortex generation would happen quicker, meaning higher $a$, as observed in the obstacle strength dependence of $\{v_c,a\}$, but which is opposite with the obstacle size.

Finally, we discuss the 2D universality of our measurement results. If the superfluid system truly possesses the 2D nature for the vortex shedding dynamics, we may ask that the dimensionless quantities, $v_c/c_s$ and $a/\xi$, should be expressed as universal functions of the dimensionless obstacle parameters, $V_0/\mu$ and $\sigma/\xi$. In Ref.~\cite{Kwon15-vd}, a single measurement point of $\{v_c,a\}$ was reported with $^{23}$Na BECs, where $v_c/c_s\approx 0.28$ and $a\xi\approx 0.11$ for $V_0/\mu\approx0.74$ and $\sigma/\xi\approx20$. We find that this value of $a\xi$ is approximately twice larger than that estimated from our measurement data for the corresponding obstacle condition. This hints that the vortex shedding dynamics in the highly oblate BEC may involve three-dimensional (3D) effects that are controlled by the condensate thickness. In the $^{23}$Na experiment, the scaled thickness of the condensate was $R_z/\xi\approx 5.5$, which is about two times thinner than the current sample.

\subsection{Critical velocity of oscillating obstacle}

We extend our vortex shedding experiment to a case with oscillating obstacle. Energy dissipation from an oscillating obstacle has been considered in many past studies with superfluid helium~\cite{Jager95,Bradley00,Niemetz02,Yano05,Sheshin08} as well as atomic superfluid gases~\cite{Raman99,Desbuquois12,Weimer15}. Being under acceleration, the oscillating obstacle provides a different boundary condition from that for a uniformly moving obstacle, thus, possibly leading to a different type of energy dissipation such as phonon or soliton emission in addition to vortex shedding~\cite{Jackson00,Radouani04,Fujimoto11,Reeves12,Khamis13}. Of particular interest is a situation where the obstacle moves with $v> v_c$, experiencing a drag force, but the oscillation amplitude is not large enough to create vortices, and a question immediately follows of how the energy accumulated around the obstacle will be resolved~\cite{Jackson00}. One might think that the energy would be resorbed by the obstacle when it turns around the end point or dissipate via phonon emission into the superfluid, stimulated by the obstacle's acceleration~\cite{Desbuquois12,Weimer15,Singh16}. To address this question based on our measurement results of $\{v_c,a\}$, we carried out a modified stirring experiment, where a penetrable obstacle sinusoidally oscillates at the center region of the BEC for 5~s. We employed our smallest obstacle with $\sigma/\xi\approx 27$ and set the oscillation amplitude to be $A\approx 45~\mu\textrm{m}$. The maximum obstacle velocity $v_m$ was controlled with the oscillation frequency $f_\textrm{osc}$ as $v_m=2\pi A f_\textrm{osc}$.

Figure 5(a) shows a series of absorption images of the stirred BEC for various $f_\textrm{osc}$, where the obstacle strength was $V_0/\mu=0.58$. As the oscillation frequency increases over a certain value of $f_{c1}\approx 2.8~\textrm{Hz}$, vortices appear in the BEC but noticeably, only in the boundary region. At the critical point, the maximum obstacle velocity is estimated to be $v_{m,c1}=2\pi A f_{c1}\approx 0.79~\textrm{mm/s}$, which we find qualitatively consistent with the critical velocity $v_c=0.74(12)~\textrm{mm/s}$ measured for the same obstacle condition (figure~4(a)). This observation suggests that phonon (or sound wave) excitations may be created by the oscillating obstacle as $v_m$ exceeds $v_c$, and transformed into vortices in the low-density boundary region of the BEC~\cite{Reeves12}. In the experiment, we observed that the vortex population was significantly suppressed with a shorter oscillation time, which implies that energy build-up is necessary for the phonon-to-vortex transformation. When the obstacle's oscillation frequency is further increased over $f_{c2}\approx 3.2~\textrm{Hz}$, we observe another threshold behavior where vortices suddenly start occupying the central region of the BEC. We understand it as a result of direct vortex generation from the obstacle for higher $v_m$. If the oscillating motion is approximated as repetitions of a pulsed linear motion with constant velocity $v_m$ over an effective distance $L_\textrm{eff} \propto A$, the direct vortex shedding requires $v_m>v_{c,L_\textrm{eff}}$, which gives the upper critical frequency $f_{c2}=\frac{v_{c,L_\textrm{eff}}}{2\pi A}=\frac{a L_\textrm{eff}}{a L_\textrm{eff}-1} \frac{v_c}{2\pi A}$. From the measured values of $\{v_c,a\}$, $f_{c2}\approx 3.2~\textrm{Hz}$ suggests $L_\textrm{eff}\approx 0.9A$.

%%%%% figure 6

\begin{figure*}[t]
 \includegraphics[width=105mm]{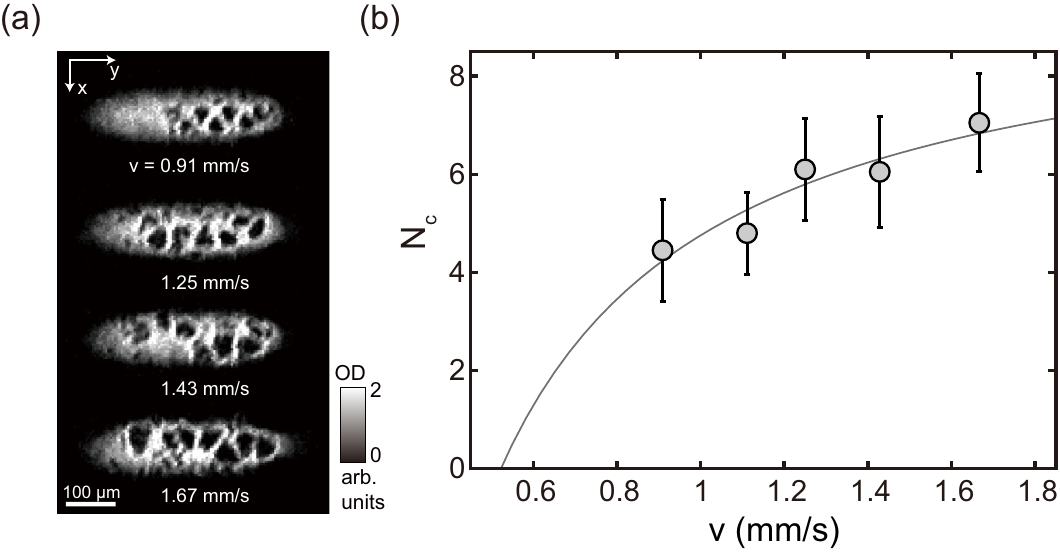}
 \centering
  \caption{Vortex cluster shedding from an impenetrable obstacle. (a) Absorption images for various $v$ with $V_0/\mu\approx 2$, $\sigma/\xi \approx 27$, and $L\approx 200~\mu$m. The charge of the vortex cluster, indicated by the area size of the density-depleted region, was increased with $v$ increasing. (b) Vortex cluster number $N_c$ as a function of the obstacle velocity $v$. Each data point was obtained from 20 measurements and its error bar indicate their standard deviation. The gray line shows a model curve fit of $N_c=2aL (1-\frac{v_c}{v})$ with $v_c=0.32 c_s$ and $a=23~\textrm{mm}^{-1}$.}
\end{figure*}

%%%%%%%%%%%

In figure~5(c), we present our whole observation results for various obstacle strengths in the plane of $V_0/\mu$ and $f_\text{osc}$, where the open circles denote the parameter positions where vortices were observed only in the boundary region, the solid circles denote those with vortices also in the central region, and the crosses denote those without vortices. It is clearly demonstrated that the energy dissipation from the oscillating obstacle develops in a two-step manner with $f_\textrm{osc}$ increasing, first only via phonon emission and then including vortex generation. The lower critical frequency $f_{c1}$, corresponding to the onset of energy dissipation, can be estimated from $v_m=v_c$, and we obtain $f_{c1}=\frac{v_c}{2\pi A}=\frac{1}{2\pi A}(1- k (V_0/\mu)^{\alpha})$ adopting the model curve of $v_c$. Following the pulsed linear motion approximation, the upper critical frequency $f_{c2}$ for direct vortex generation can be expressed as $f_{c2}=\frac{a L_\textrm{eff}}{a L_\textrm{eff}-1} f_{c1}$. We find the estimated $f_{c1,c2}$ in reasonable agreement with the experimental results (figure~5(c)), with $L_\textrm{eff}=0.9 A$. For weak obstacles with $V_0/\mu<0.4$, the observed $f_{c1}$ slightly deviates upwards from the prediction, which we attribute to the retardation of the phonon-to-vortex transformation due to weak perturbations. To sum up the results, we conclude that the periodic shedding model, which parametrizes the dissipation processes with $\{v_c, a\}$, effectively describes the responses of the superfluid to the oscillation of the penetrable obstacle, even including the phonon emission regime.

It is noteworthy that for the weakest obstacle with $V_0/\mu\approx0.16$, indeed, we could not perform the $f_v$ measurement with the linear sweeping method due to low $a$ as well as short $L$. From the extrapolation of the results in figure 4(a), the value of $a$ is estimated to be about $50~\textrm{mm}^{-1}$, giving $a L_\textrm{eff}\approx 2.0$ and thus making the direct shedding of vortex dipoles difficult. With the oscillating method, however, we could detect the onset of energy dissipation thanks to the indirect vortex generation at the sample boundary. This shows that using an oscillating obstacle can provide an efficient way to determine $v_c$ once the population growth of phonon excitations can be reliably and immediately measured~\cite{Raman99,Desbuquois12,Weimer15}.

\subsection{Asymptotic relation of $a$ to $\textrm{St}$}

The Strouhal number $\textrm{St}$ is defined in the impenetrable regime but is also associated with the shedding frequency. Thus, it is intriguing to discuss a possible relation between $a$ and $\textrm{St}$ as $V_0/\mu$ increases into the impenetrable regime. Assuming that the characteristic relation $f_v=a(v-v_c)$ still holds in the impenetrable regime~\cite{Frisch92,Winiecki99,Winiecki00}, the Strouhal number is given by
\begin{equation}
    \textrm{St} = f_v D/v = aD(1-\frac{v_c}{v}).
\end{equation}
This predicts the saturation behavior of $\textrm{St}\rightarrow \textrm{St}_\infty=aD$ for a sufficiently fast obstacle with $v \gg v_c$, as observed in previous works. If the saturation value $\textrm{St}_\infty$ is a universal constant independent of the obstacle diameter $D$, it implies the $1/D$ dependence of $a$, which seems to be in accordance with our observation of the decrease of $a$ with increasing $\sigma$ (figure 4(b)), although the obstacle diameter is ill-defined in the penetrable regime. The inset of figure 4(b) shows $a\sigma$ as function of $\sigma/\xi$. Note that the diameter of a Gaussian optical obstacle is typically calculated as $D=\sigma \sqrt{2 \ln \frac{V_0}{\mu}}$, that is the diameter of the density-depleted hole induced by a stationary obstacle. 

In figure 6, we present a set of vortex shedding data for an impenetrable obstacle with $V_0/\mu\approx2$ and $\sigma/\xi\approx 27$. We linearly translate the impenetrable obstacle for the distance $L\approx200~\mu\textrm{m}$ and measure the cluster number $N_c$ for various obstacle velocities. As the velocity exceeds a certain threshold value, we observe the regular shedding of the vortex clusters and $N_c$ is increased up to about 8 for our experimental range of $v$. With the assumption of $f_v=a(v-v_c)$ in the impenetrable regime, the cluster number is given by $N_c=2f_v \frac{L}{v}=2aL(1-\frac{v_c}{v})$, where the factor of 2 accounts for that one shedding cycle contains two cluster emissions with opposite net circulation. From a fit of the model curve to the measurement results of $N_c$, we obtain $v_c\approx 0.32 c_s$ and $a\approx 23~\textrm{mm}^{-1}$. In comparison to the measured values of $\{v_c, a\}$ at $V_0/\mu \approx 1$, the critical velocity is almost maintained, whereas the $a$ value is reduced by a factor 10. This corroborates the observed behavior of $v_c$ and $a$ as $V_0/\mu$ increases over unity in figure 4(a). The measured value of $a$ suggests $\textrm{St}_\infty\approx0.16$, which is remarkably similar with the prediction of $\textrm{St}\approx0.14$ from the numerical study~\cite{Reeves15}.

\section{Summary}

We investigated the periodic vortex shedding from a penetrable obstacle in a BEC, and measured the shedding frequency $f_v$ for a wide range of the obstacle parameters including velocity, potential strength, and size. We confirmed that the shedding frequency is well characterized by the linear relationship of  $f_v=a(v-v_c)$, and investigated the characteristic dependence of $\{v_c, a\}$ on the obstacle strength and size. We also investigated the critical vortex shedding dynamics of an oscillating obstacle and demonstrated that the energy dissipation from the obstacle develops in a two-step manner with the oscillation frequency increasing, starting from phonon emission and then including vortex generation. The upper and lower critical frequencies $f_{c1,c2}$ were found to be well explained by the periodic shedding model and furthermore, in good quantitative agreement with the measured values of $\{v_c, a\}$. Finally, we discussed the asymptotic relation of $a$ to the Strouhal number in the impenetrable regime. We expect that the measurement results provided in this work will be beneficial to establish the phenomenological understanding of the universal shedding dynamics of superfluid.

\begin{acknowledgments}
This work was supported by the National Research Foundation of Korea (NRF-2018R1A2B3003373, NRF-2019M3E4A1080400) and the Institute for Basic Science in Korea (IBS-R009-D1).
\end{acknowledgments}

\end{document}